\documentclass[aps,prl,twocolumn,showpacs,amsmath,amssymb,amsfonts,nofootinbib,long,preprintnumbers]
{revtex4}
\usepackage{epsfig,latexsym,bm}

\newcommand\keV{\mbox{keV}}

\newcommand\GeV{\mbox{GeV}}
\newcommand\TeV{\mbox{TeV}}

\newcommand\D{\partial}

\begin{document}

\title{Spinning Supersymmetric Q-balls}

\author{L. Campanelli$^{1,2}$}
\email{leonardo.campanelli@ba.infn.it}
\author{M. Ruggieri$^{1,2}$}
\email{marco.ruggieri@ba.infn.it}

\affiliation{$^1$Dipartimento di Fisica, Universit\`{a} di Bari, I-70126 Bari, Italy}
\affiliation{$^2$INFN - Sezione di Bari, I-70126 Bari, Italy}

\date{April, 2009}


\begin{abstract}
We construct nontopological solitonic solutions in (3+1)-dimensional Minkowski spacetime carrying a conserved global $U(1)$ charge and nonvanishing angular momentum in a supersymmetric extension of the standard model with low-energy, gauge-mediated symmetry breaking.
\end{abstract}


\maketitle
\preprint{BA-TH/610-09}



\section{I. Introduction}

Q-balls are nontopological solitonic solutions of a self-interacting complex scalar field theory carrying a conserved global $U(1)$ charge. Introduced by Coleman in 1985~\cite{Coleman}, their properties have been extensively studied since then~\cite{Q-balls}.

If on the one hand, Volkov and W\"{o}hnert~\cite{Volkov} showed, in the context of some theories with nonrenormalizable scalar potentials, that there exist particular Q-ball configurations possessing nonvanishing angular momentum, now known as ``spinning Q-balls'', on the other hand, Dvali, Kusenko and Shaposhnikov~\cite{Dvali} proved in the framework of supersymmetric extensions of the standard model that gauge-singlet combinations of squarks and sleptons corresponding to some flat direction of the supersymmetric potential can give rise to Q-balls whose charge $Q$ is some combination of baryon and lepton numbers. The attractive feature of theses ``supersymmetric Q-balls'' is that they could represent the dark matter component of the universe~\cite{Kusenko-Shaposhnikov,KusenkoTalk} (for reviews on dark matter see, e.g., Ref.~\cite{DarkMatter}). Motivated by this fact, experimental searches for Q-balls are being carried out~\cite{Experiment}, although no compelling evidence for their existence has been reported so far.

The aim of this paper is to present novel configurations of a charged scalar field describing nonspherically-symmetric supersymmetric Q-balls with nonvanishing angular momentum. To our knowledge, {\it spinning supersymmetric Q-balls} are the first example of analytical solution in field theory in Minkowski spacetime representing a soliton possessing angular momentum. As we will see, spinning supersymmetric Q-balls are excitations of spherically-symmetric supersymmetric Q-balls, since their energy spectrum lies above the ground state represented by nonspinning Q-balls. However, it is highly probable that these exited states could form during collisions between supersymmetric Q-balls and, most importantly, during the process of fragmentation of the Affleck-Dine condensate, which is a plausible process that can lead to a copious production of Q-balls at the end of inflation~\cite{Kusenko-Shaposhnikov,KusenkoTalk}.

The plan of the paper is as follows. In Section II we review the general properties of spherically-symmetric supersymmetric Q-ball in order to make clearer the derivation and the study of nonspherically-symmetric supersymmetric Q-balls configurations which will be tackled in Section III. In Section IV we show that spinning supersymmetric Q-balls are stable against small perturbations about their classical configurations. Finally, in Section V we draw our conclusions.


\section{II. Spherically-symmetric Supersymmetric Q-balls}

Let us consider a charged scalar field $\Phi$ whose dynamics is described by lagrangian density
\begin{equation}
\label{eq:Lagr1} \mathcal{L} = (\D_\mu \Phi^*)(\D^\mu \Phi) - U(|\Phi|).
\end{equation}
%
%
Being the theory invariant under a global $U(1)$ transformation, there exists a conserved Noether charge, $q$, which we normalize as
\begin{equation}
\label{eq:charge} q = \frac{1}{i} \int \! d^3x \left(\Phi^* \dot{\Phi} - \Phi \dot{\Phi}^* \right) \!,
\end{equation}
where a dot indicates a derivative with respect to time. The energy-momentum tensor associated to a given field configuration $\Phi(t,\textbf{r})$, reads
\begin{equation}
\label{energy-momentum} T^\mu_\nu = \D^\mu \Phi \D_\nu \Phi^* + \D_\nu \Phi \D^\mu \Phi^* -\delta^\mu_\nu \mathcal{L},
\end{equation}
so that the total energy is
\begin{equation}
\label{eq:energy} E \equiv \int \! d^3x \, T_0^0 = \int \! d^3x \left[ |\dot{\Phi}|^2 + |{\nabla \Phi}|^2 + U(|\Phi|) \right] \!.
\end{equation}
In general, Q-balls are solitonic solutions of the field equations carrying a definite value of the charge~\eqref{eq:charge}, let us say $Q$. An elegant way to construct such a type of solution~\cite{Kusenko} is to introduce a Lagrange multiplier $\omega$ associated to $q$, and require that the physical configuration $\Phi(t,\textbf{r})$ makes the functional
\begin{equation}
\label{functional1} \mathcal{E}[\Phi,\Phi^*,\omega] \equiv E + \omega \left[Q - \frac{1}{i} \int \! d^3x \left(\Phi^* \dot{\Phi} - \Phi \dot{\Phi}^* \right) \right]
\end{equation}
stationary with respect to independent variations of $\Phi$ and $\omega$. Noticing that the choice~\cite{Coleman,Kusenko}
\begin{equation}
\label{Phit} \Phi(t,\textbf{r}) = e^{i\omega t} \phi(\textbf{r}),
\end{equation}
assures that the total energy $E$ is independent on the time, one finds that Q-balls solutions have to satisfy the constraints
\begin{equation}
\label{eq:PhysCond} \frac{\delta \mathcal{E}}{\delta\phi} = 0, \;\;\;\; \frac{\delta \mathcal{E}}{\delta\phi^*} = 0, \;\;\;\; \frac{\delta \mathcal{E}}{\delta\omega} = 0.
\end{equation}
The first two constraints lead to the equations of motion of the fields $\phi^*(\textbf{r})$ and $\phi(\textbf{r})$, respectively,
\begin{eqnarray}
\label{eq:eqMot3bis} && \left( \nabla^2 + \omega^2 \right) \! \phi^* = \frac{\delta U}{\delta \phi} \, ,
\\
\label{eq:eqMot3} && \left( \nabla^2 + \omega^2 \right) \! \phi = \frac{\delta U}{\delta \phi^*} \, ,
\end{eqnarray}
while the second one is equivalent to the requirement that the charge corresponding to the solution of the equation of motion is equal to $Q$:
\begin{equation}
\label{q=Q} q = 2\omega \! \int \! d^3x \, |\phi|^2 \equiv Q.
\end{equation}
Taking into account the equations of motion, the functional $\mathcal{E}$ can be conveniently re-written as:
\begin{equation}
\label{FunctionalE} \mathcal{E} = \int \! d^3x \! \left[ U - \phi \, \frac{\delta U}{\delta \phi} \right] + \omega Q.
\end{equation}
A spherically-symmetric Q-ball is defined as the solution $\phi(r)$ of Eq.~\eqref{eq:eqMot3} satisfying, at fixed charge $Q$, the boundary conditions~\cite{Coleman}
\begin{equation}
\label{boundary} \lim_{r \rightarrow 0}\phi(r) \neq \infty, \;\;\; \lim_{r \rightarrow \infty} \phi(r) = 0,
\end{equation}
where $r \equiv |\textbf{r}|$. In particular, a (spherically-symmetric) supersymmetric Q-ball is a Q-ball configuration arising in a supersymmetric model of particle physics where supersymmetry is broken via low-energy gauge mediation~\cite{SUSY}. In this kind of model the coupling of the massive vector-like messenger fields to the gauge multiplets, with coupling constant $g\sim 10^{-2}$, leads to the breaking of supersymmetry~\cite{SUSY}. The coupling itself gives rise to an effective potential for the flat direction $\phi$ whose lowest order (two-loops) contribution has been calculated in Ref.~\cite{de Gouvea}:
\begin{equation}
\label{potential} U(z) = \Lambda \! \int_0^1 \!\! dx \, \frac{z^{-2} - x(1-x) + x(1-x)\ln[x(1-x)z^2]}{[z^{-2}-x(1-x)]^2} \, .
\end{equation}
Here, $z \equiv |\phi|/M$ and $M \equiv M_S/(2g)$, with $M_S$ the messenger mass scale. The value of the mass parameter $\Lambda^{1/4}$ is constrained as (see, e.g., Ref.~\cite{Kasuya511}):
\begin{equation}
\label{LambdaLimit} 1 \TeV \, \lesssim \Lambda^{1/4} \lesssim \, 10^4 \! \left( \frac{m_{3/2}}{\GeV} \right)^{\!1/2} \TeV,
\end{equation}
where the gravitino mass, $m_{3/2}$, is in the range $100 \, \keV \lesssim m_{3/2} \lesssim 1\GeV$~\cite{de Gouvea,Kasuya511}. The asymptotic expressions of $U(z)$, for small and large $z$ are~\cite{de Gouvea}:
\begin{equation}
\label{potentialapprox} \frac{U(z)}{\Lambda} \simeq
    \left\{ \begin{array}{ll}
        z^2, &  \;\; \mbox{if} \;\; z \ll 1, \\
        (\ln z^2)^2 - 2 \ln z^2 + \frac{\pi^2}{3} \, , &  \;\; \mbox{if} \;\; z \gg 1.
    \end{array}
    \right.
\end{equation}
A widely used approximation in constructing Q-ball solutions, whose validity has been ascertained in Ref.~\cite{Campanelli-Ruggieri}, consists in replacing the full potential $U(z)$ with its asymptotic expansions (\ref{potentialapprox}) in which a plateau plays the role of the logarithmic rise for large values of $z$. More precisely, the approximate supersymmetric potential has the form
\begin{equation}
\label{Uapproximate} U(|\phi|) \equiv
    \left\{ \begin{array}{ll}
        m_\phi^2 |\phi|^2, &  \;\; \mbox{if} \;\; |\phi| \leq M, \\
        \Lambda \, , &  \;\; \mbox{if} \;\; |\phi| \geq M,
    \end{array}
    \right.
\end{equation}
where $m_\phi \equiv \sqrt{\Lambda}/M$ is the soft breaking mass and is of order $1\TeV$~\cite{Kasuya511}. Within this approximation, it has been shown that the potential $U(|\phi|)$ allows spherically-symmetric Q-ball solutions as the nonperturbative ground state of the model~\cite{Dvali,Kusenko-Loveridge}.

The profile of the supersymmetric Q-ball is easily found from Eqs.~\eqref{eq:eqMot3} and \eqref{Uapproximate}:
\begin{equation}
\label{profile} \phi(r) =
    \left\{ \begin{array}{ll}
        \phi_0 j_0(\omega r)
        + \tilde{\phi}_0 y_0(\omega r), & \;\; \mbox{if} \;\; |\phi| \geq M,
        \\ \\
        \phi_\delta \, i_0(r/\delta)
        + \tilde{\phi}_\delta k_0(r/\delta), &  \;\; \mbox{if} \;\; |\phi| \leq M,
    \end{array}
    \right.
\end{equation}
where $\phi_0$ and $\phi_\delta$ are constants of integration, $j_0(x) = \sin x/x$ is the zeroth-order spherical Bessel function of first kind, $y_0(x) = -\cos x/x$ is the zeroth-order spherical Bessel function of second kind, $i_0(x) = \sinh x/x$ is the zeroth-order modified spherical Bessel function of first kind, and $k_0(x) = e^{-x}/x$ is the zeroth-order modified spherical Bessel function of second kind~\cite{Abramowitz}. Here, we have introduced the ``thickness'' of the Q-ball,
\begin{equation}
\label{delta} \delta \equiv \frac{1}{\sqrt{m_\phi^2 - \omega^2}} \, .
\end{equation}
We can now define the ``radius'' of the Q-ball, $R$, as the solution of the equation $j_0(\omega R) = M/\phi_0$. In the limit $\phi_0 \gg M$ we have $j_0(\omega R) \simeq 0$, from which it follows that
\begin{equation}
\label{RadiusOmega} R = \frac{\pi}{\omega} \, .
\end{equation}
(We will se in the following that the condition $\phi_0 \gg M$ will correspond to have large values of the charge $Q$.) If the thickness of the Q-ball is much smaller than its radius (we will see, below, that indeed large charges $Q$ implies that $R \gg \delta$), we can write
\begin{equation}
\label{profilebis} \phi(r) =
    \left\{ \begin{array}{ll}
        \phi_0 j_0(\omega r), &  \;\; \mbox{if} \;\; r \leq R,
        \\ \\
        0 , &  \;\; \mbox{if} \;\; r \geq R,
    \end{array}
    \right.
\end{equation}
which is the solution found in Ref.~\cite{Dvali,Kusenko-Loveridge}. Inserting the above solution in Eq.~\eqref{FunctionalE} and minimizing with respect to $\omega$ [see last equation in Eq.~\eqref{eq:PhysCond}], we find the parameter $\omega$ as a function of the charge $Q$:
\begin{equation}
\label{omegaL} \frac{\omega}{m_\phi} = 2\sqrt{2} \, \pi \! \left(\frac{Q}{Q^{\rm (cr)}}\right)^{\! -1/4} \! ,
\end{equation}
where we have introduced the ``critical charge'' $Q^{\rm (cr)}$, whose meaning will be clear in the following, as
\begin{equation}
\label{Qcr1} Q^{\rm (cr)} \equiv \frac{4\Lambda}{m_\phi^4} \, .
\end{equation}
Inserting Eq.~\eqref{omegaL} in Eq.~\eqref{RadiusOmega}, we find the Q-ball radius as a function of the charge:
\begin{equation}
\label{raggioL} \frac{R}{m_\phi^{-1}} = \frac{1}{2\sqrt{2}} \left(\frac{Q}{Q^{\rm (cr)}}\right)^{\! 1/4} \! .
\end{equation}
From the above equation and taking into account Eqs.~\eqref{delta} and~\eqref{omegaL}, we find that for large charges, $Q \gg Q^{\rm(cr)}$, it results $R/\delta \simeq (Q/Q^{\rm (cr)})^{1/4}/2\sqrt{2} \gg 1$, and this justifies our approximation to neglect the thickness of the Q-ball in computing its profile.

Inserting Eq.~\eqref{omegaL} in Eq.~\eqref{FunctionalE}, and observing that, at fixed charge $Q$, the energy coincides with the functional $\mathcal{E}$, we find $E$ as a function of the charge:
\begin{equation}
\label{EnergyL} \frac{E}{m_\phi Q_{\rm cr}} = \frac{4\pi}{3} \left(\frac{Q}{Q^{\rm (cr)}}\right)^{\! 3/4} \! .
\end{equation}
Finally, inserting Eq.~\eqref{profilebis} in Eq.~\eqref{q=Q} and taking into account Eq.~\eqref{omegaL}, we find the value of $\phi_0$ as a function of the charge:
\begin{equation}
\label{phi0} \frac{\phi_0}{M} = \left(\frac{Q}{Q^{\rm (cr)}}\right)^{\! 1/4} \! .
\end{equation}
The above relation clarifies the meaning of the critical charge: The Q-ball solution we found in the limit $\phi_0 \gg M$, corresponds indeed to the case of large charges compared to $Q^{\rm (cr)}$.

If the energy $E$ of the Q-ball at fixed charge $Q$ is less then $m_\phi Q$, the soliton decays into $Q$ quanta of the field (the perturbative spectrum of the theory), each of them with mass $m_\phi$. Instead, if $E < m_\phi Q$ the Q-ball is said to be classically stable, and then represents the ground state of the theory. Using Eq.~\eqref{EnergyL}, we find classical stability, $E/m_\phi Q < 1$, for $Q > Q^{(\rm min)}$, with
\begin{equation}
\label{Qmin0} Q^{(\rm min)} = \left( \frac{4\pi}{3} \right)^{\!\!4} Q^{(\rm cr)}.
\end{equation}


\section{III. Spinning Supersymmetric Q-balls}

In general, spherically-symmetric Q-balls have zero angular momentum. In fact, the total angular momentum for a scalar field configuration is given by
\begin{equation}
\label{J}  \textbf{J} = (J^{23},J^{31},J^{12}),
\end{equation}
where $J^{\mu \nu}$ is the total angular momentum tensor~\cite{Bjorken}
\begin{equation}
\label{JJ} J^{\mu \nu} = \int \! d^3x \left( x^{\mu} T^{0\nu} -x^{\nu} T^{0\mu} \right),
\end{equation}
with $T^{\mu\nu}$ being the energy-momentum tensor given by Eq.~\eqref{energy-momentum}. Now, using spherical coordinates, $\textbf{r} = (r,\theta,\varphi)$, we obtain
\begin{eqnarray}
\label{J1} && \!\!\!\!\!\!\!\!\!\!  J^{1} = \int \! d^3x \! \left[ \cos(2\theta) \sin \varphi \, T^0_\theta + \cot \! \theta \cos \varphi \, T^0_\varphi \right] \! ,
\\
\label{J2} && \!\!\!\!\!\!\!\!\!\! J^{2} = \int \! d^3x \! \left[ -\cos(2\theta) \cos \varphi \, T^0_\theta + \cot \! \theta \sin \varphi \, T^0_\varphi \right] \! ,
\\
\label{J3} && \!\!\!\!\!\!\!\!\!\! J^{3} = - \int \! d^3x \, T^0_\varphi \, ,
\end{eqnarray}
so that for a spherically-symmetric Q-ball, $\Phi = \Phi(t,r)$, we get $\textbf{J} = 0$. On the other hand, for a nonspherically-symmetric Q-ball (if it ever exists), $\Phi = \Phi(t,r,\theta,\varphi)$, we could have in principle a nonvanishing angular momentum (for supersymmetric Q-balls this will be indeed the case). In particular, if one makes use of the ``axially-symmetric ansatz'',
\begin{equation}
\label{ansatz} \Phi(t,r,\theta,\varphi) = \Phi(t,r,\theta) \, e^{im\varphi},
\end{equation}
$m$ being a real constant, one easily finds
\begin{equation}
\label{J3Q} J^3 = - m q,
\end{equation}
where $q$ is given by Eq.~\eqref{eq:charge}. Since single-valuedness of the scalar field requires $\Phi(t,r,\theta,\varphi + 2\pi) = \Phi(t,r,\theta,\varphi)$, the constant $m$ must be an integer. Therefore, for this particular configuration, the third component of the angular momentum is quantized and proportional to the charge. It is useful for the following discussion to observe that if a Q-ball configuration is such that the field $\Phi$ is given by
\begin{equation}
\label{ansatzbis} \Phi(t,r,\theta,\varphi) = e^{i\omega t + im\varphi} \phi(r,\theta),
\end{equation}
with $\phi(r,\theta)$ a real function, then (as it easy to verify) it results $J^1 = J^2 = 0$, so that the nonspherically-symmetric Q-ball is indeed a spinning Q-ball with total angular momentum directed along the $z$-axis.

We now return to the supersymmetric case to find nonspherically-symmetric supersymmetric Q-ball configurations. We start by writing the equation of motion for the field $\phi(\textbf{r})$ [Eq.~\eqref{eq:eqMot3}] in spherical coordinates:
\begin{equation}
\label{motion} \left( \! \D_r^2 + \frac{2}{r} \, \D_r  + \frac{1}{r^2} \, \D^2_\theta + \frac{\cot \! \theta}{r^2} \, \D_\theta + \frac{\csc^2 \! \theta}{r^2} \, \D^2_\varphi + \omega^2 \! \right) \!\! \phi = \frac{\delta U}{\delta \phi^*} .
\end{equation}
Using the technique of separation of variables,
\begin{equation}
\label{separation} \phi(r,\theta,\varphi) \equiv \phi(r) \, \phi(\theta,\varphi),
\end{equation}
we easily find [using the approximate form of the supersymmetric potential, Eq.~\eqref{Uapproximate}] the solution of Eq.~\eqref{motion}:
\begin{equation}
\label{thetavarphi} \phi(\theta,\varphi) = \sqrt{4\pi} Y_l^m (\theta,\varphi),
\end{equation}
where
\begin{equation}
\label{harmonics} Y_l^m (\theta,\varphi) \equiv \sqrt{\frac{2l+1}{4\pi} \, \frac{(l-m)!}{(l+m)!}} \: P_l^m(\cos \theta) \, e^{im\varphi}
\end{equation}
are the usual spherical harmonics of degree $l$ and order $m$, with $P_l^m(x)$ being the associated Legendre polynomials of degree $l$ and order $m$~\cite{Abramowitz}, and
\begin{equation}
\label{profileS} \phi(r) =
    \left\{ \begin{array}{ll}
        \phi_l j_l(\omega r)
        + \tilde{\phi_l} y_l(\omega r), &  \;\; \mbox{if} \;\; |\phi| \geq M,
        \\ \\
        \phi_\delta \, i_l(r/\delta)
        + \tilde{\phi}_\delta k_l(r/\delta), &  \;\; \mbox{if} \;\; |\phi| \leq M.
    \end{array}
    \right.
\end{equation}
Here, $\phi_l$, $\tilde{\phi}_l$, $\phi_\delta$, $\tilde{\phi}_\delta$ are constants of integration, the Q-ball thickness $\delta$ is the same as in Eq.~\eqref{delta}, $j_l(x)$ and $y_l(x)$ are the spherical Bessel function of order $l$ of first and second kind respectively, and $i_l(x)$ and $k_l(x)$ are the modified spherical Bessel function of order $l$ of first and second kind respectively~\cite{Abramowitz}.

We can now define the ``radius'' of the Q-ball, $R_l$, as the solution of the equation $j_l(\omega R_l) = M/\phi_l$. In the limit $\phi_l \gg M$ we have $j_l(\omega R_l) \simeq 0$, from which it follows that
\begin{equation}
\label{RadiusOmegal} R_l = \frac{j_{l+1/2,1}}{\omega} \, ,
\end{equation}
where $j_{\nu,1}$ represents the first zero of the Bessel function of order $\nu$ of first kind, $J_\nu(x)$~\cite{Abramowitz}.
\footnote{We observe that $j_{\nu,1}$ is an increasing function of $\nu$ with $j_{1/2,1} = \pi$ and $j_{3/2,1} \simeq 4.49341$, $j_{5/2,1} \simeq 5.76346$, $j_{7/2,1} \simeq 6.98793$, $j_{9/2,1} \simeq 8.18256$, $j_{11/2,1} = 9.35581$~\cite{Abramowitz}, etc. Moreover, the asymptotic expansion of $j_{\nu,1}$, as $\nu \rightarrow \infty$, is: $j_{\nu,1} \sim \nu - (a_1 / \, 2^{1/3}) \, \nu^{1/3} + (3 \,a_1^2 \,2^{1/3} \!/\,20) \, \nu^{-1/3} + ...$, where $a_1 \simeq -2.33811$ is the first negative zero of the Airy function $\mbox{Ai}(x)$~\cite{Abramowitz}.}
(We will se in the following that the condition $\phi_l \gg M$ will correspond to have large values of the charge $Q$.) If the thickness of the Q-ball is much smaller than its radius (we will see, below, that indeed large charges $Q$ implies that $R_l \gg \delta$), we can write
\begin{equation}
\label{profileSbis} \phi(r) =
    \left\{ \begin{array}{ll}
        \phi_l j_l(\omega r), &  \;\; \mbox{if} \;\; r \leq R_l,
        \\ \\
        0 , &  \;\; \mbox{if} \;\; r \geq R_l.
    \end{array}
    \right.
\end{equation}
Inserting the above solution in Eq.~\eqref{FunctionalE} and minimizing with respect to $\omega$, we find the parameter $\omega$ as a function of the charge $Q$:
\begin{equation}
\label{omegal} \omega_l = \left( \frac{j_{l+1/2,1}}{\pi} \right)^{\! 3/4} \omega_0,
\end{equation}
where, from now on, quantities with the subscript ``0'' refer to the case of spherically-symmetric supersymmetric Q-balls analyzed in Section II.

Inserting Eq.~\eqref{omegal} in Eqs.~\eqref{RadiusOmegal} and \eqref{FunctionalE} we find, respectively, the Q-ball radius and energy as a function of the charge:
\begin{equation}
\label{Rl} R_l = \left( \frac{j_{l+1/2,1}}{\pi} \right)^{\! 1/4} R_0
\end{equation}
and
\begin{equation}
\label{El} E_l = \left( \frac{j_{l+1/2,1}}{\pi} \right)^{\! 3/4} E_0,
\end{equation}
while, inserting Eq.~\eqref{profileSbis} in Eq.~\eqref{q=Q} and taking into account Eq.~\eqref{omegal}, we obtain 
\begin{equation}
\label{phil} \frac{\phi_l}{M} = \left(\frac{Q}{Q^{\rm (cr)}_l}\right)^{\! 1/4} \! ,
\end{equation}
where we have introduced the ``critical charge'' $Q^{\rm (cr)}_l$ as
\begin{equation}
\label{Qcrl} Q^{\rm (cr)}_l \equiv \pi (j_{l+1/2,1})^3 \, [j_{l+1}(j_{l+1/2,1})]^4 \, Q^{\rm (cr)}_0.
\end{equation}
%
Therefore, the Q-ball solution we found in the limit $\phi_l \gg M$, corresponds indeed to the case of large charges compared to $Q^{\rm (cr)}_l$. Also, from Eq.~\eqref{Rl} and taking into account Eqs.~\eqref{delta} and~\eqref{omegal}, we find that for large charges, $Q \gg Q^{\rm (cr)}_l$, it results $R_l/\delta \simeq R_l/m_\phi^{-1} \gg 1$, and this justifies our approximation to neglect the thickness of the Q-ball in computing its profile.

Finally, using Eq.~\eqref{El}, we find classical stability, $E_l/m_\phi Q < 1$, for $Q > Q^{(\rm min)}_l$, with
\begin{equation}
\label{Qminl} Q^{\rm (min)}_l = \left( \frac{j_{l+1/2,1}}{\pi} \right)^{\! 3} Q^{\rm (min)}_0.
\end{equation}
Taking into account that the general form of a supersymmetric Q-ball solution [given by Eqs.~\eqref{Phit}, \eqref{separation}-\eqref{harmonics}, and \eqref{profileSbis}] is of the form~\eqref{ansatzbis} with $\phi(r,\theta)$ a real function, and taking into account the discussion at the beginning of this Section, we conclude that the nonspherically-symmetric Q-ball solutions we found describe Q-balls with total angular momentum directed along the $z$-axis and equal to $J^3 = -mQ$. 
Moreover, observing that the angular part of the Q-ball solution is proportional to the spherical harmonic $Y_l^m (\theta,\varphi)$, we deduce that $l$ determines its parity $P$:
\begin{equation}
\label{Parity} P = (-1)^l.
\end{equation}
However, not all values of $l$ are admitted since, at fixed charge $Q$, a Q-ball with definite angular momentum $J^3$ (or, which is the same, with definite value of $m$) is such that its energy is minimum. Looking at Eq.~\eqref{El} and taking into account that $j_{l+1/2,1}$ is an increasing function of $l$, we deduce that at fixed $Q$ and $m$, two values of $l$ are allowed: For even (odd) $|m|$, $l = |m|$ if the parity of the Q-ball solution is positive (negative) and $l = |m| + 1$ if the parity of the Q-ball solution is negative (positive). Accordingly, the energy spectrum of allowed states of a spinning supersymmetric Q-ball 
looks like that in Fig.~1.


\begin{figure}[t]
\begin{center}
\includegraphics[clip,width=0.45\textwidth]{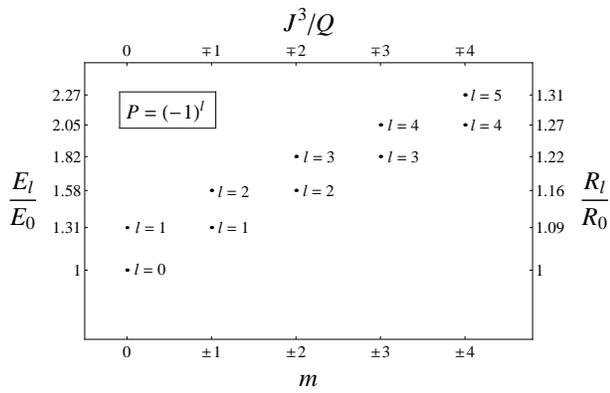}
\caption{Energy spectrum of allowed states of a spinning supersymmetric Q-ball for $-4 \leq m \leq 4$. Also shown are the corresponding values of the radius $R_l$, the angular momentum $J^3$, and the parity $P$ of the state.}
\end{center}
\end{figure}


In Fig.~2, we plot the spinning supersymmetric Q-ball's profile, $\phi(r,\theta) \equiv \phi(r,\theta,\varphi) \, e^{-im\varphi}$, as a function of $r$ and $\theta$ for different values of $l$ and $m$, at fixed charge $Q = 5 \times 10^2 Q^{\rm (min)}_l$. We observe that, using well-known properties of spherical harmonics, the profiles for negative values of $m$ coincide with the corresponding positive ones if $|m|$ is even, while they get an extra minus sign if $|m|$ is odd.


\begin{figure}[h!]
\begin{center}

\includegraphics[clip,width=0.45\textwidth]{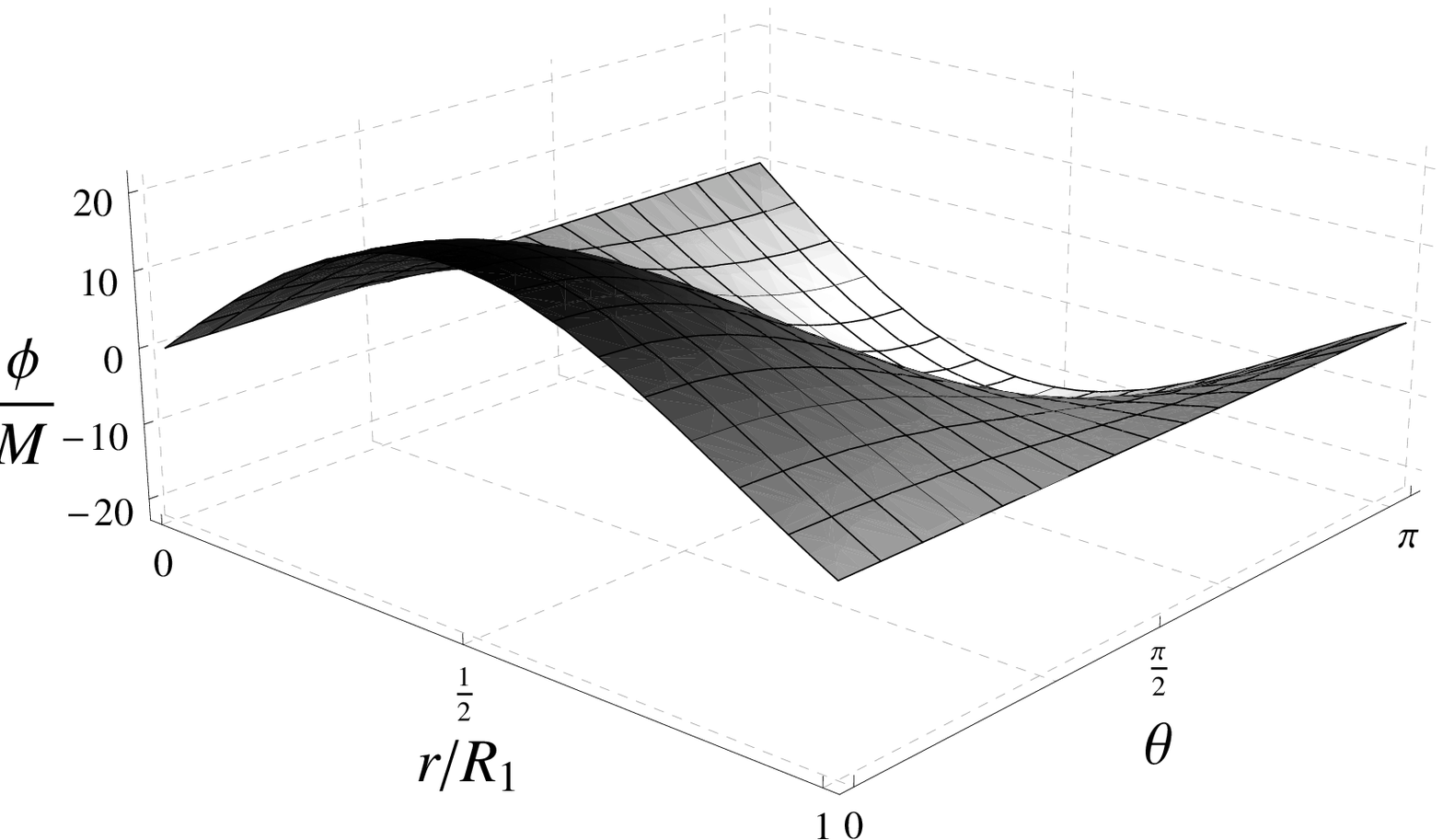}

\vspace*{0.3cm}

\includegraphics[clip,width=0.45\textwidth]{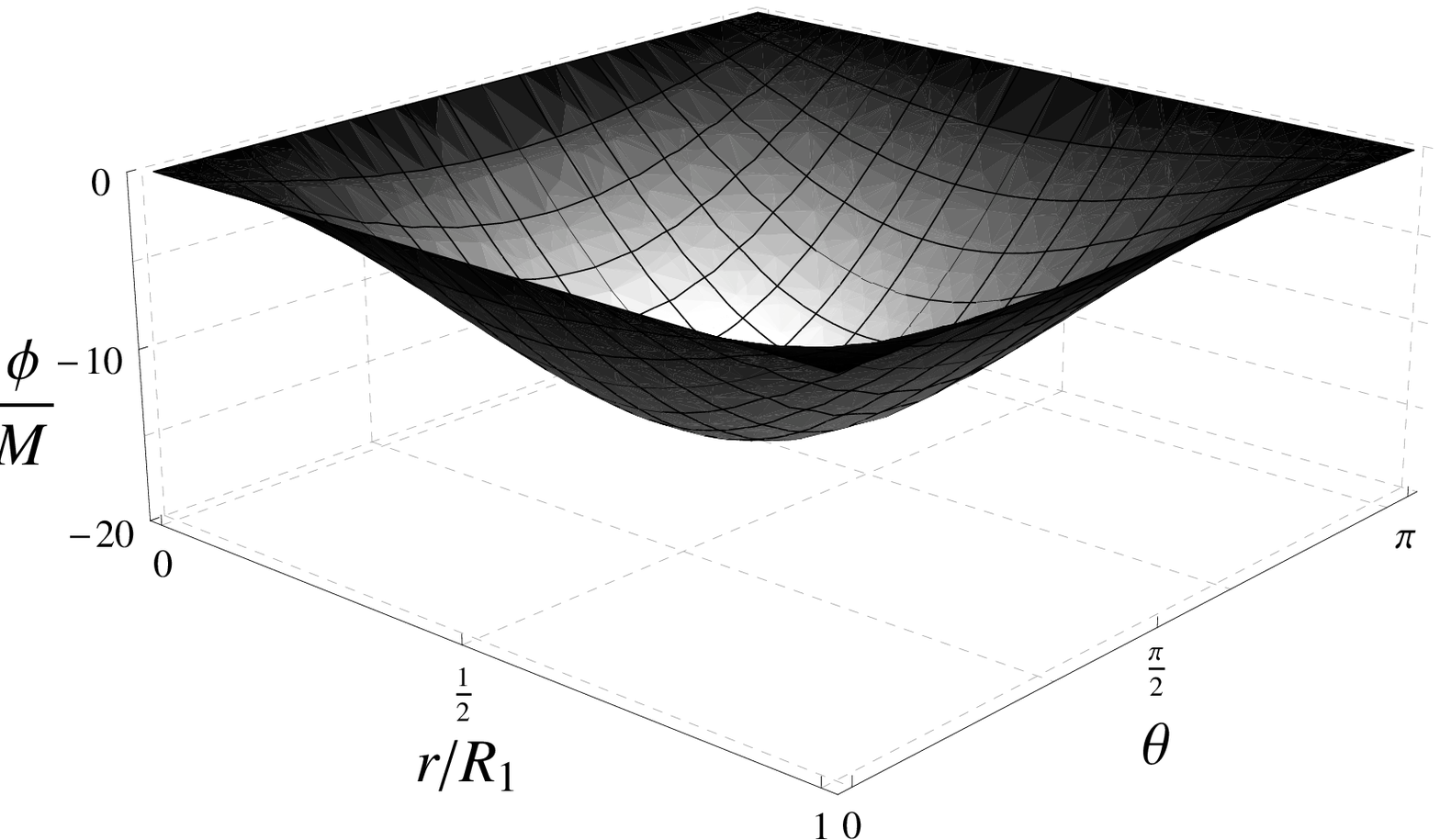}

\vspace*{0.3cm}

\includegraphics[clip,width=0.45\textwidth]{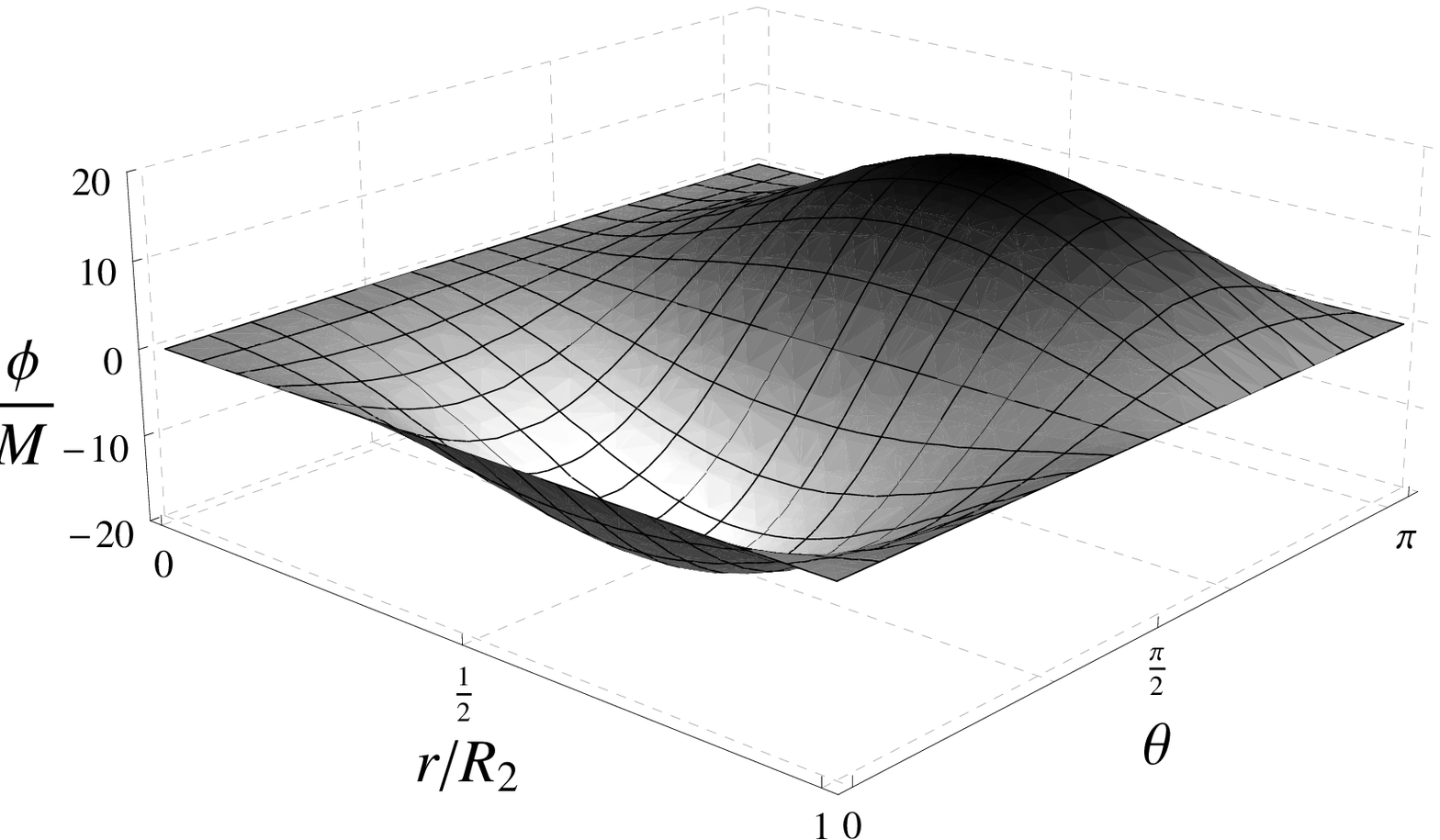}

\caption{Spinning supersymmetric Q-ball's profile, $\phi(r,\theta) \equiv \phi(r,\theta,\varphi) \, e^{-im\varphi}$, for $Q = 5 \times 10^2 Q^{\rm (min)}_l$. From upper to lower panel: $(l,m)=(1,0)$, $(1,1)$, $(2,1)$.}
\end{center}
\end{figure}


\section{IV. Stability}

Before concluding, we would like to show that spinning supersymmetric Q-balls are stable against small perturbations about their classical configurations. We will closely follow an analysis performed in Ref.~\cite{Radu}  (see also references therein) regarding the stability of Q-balls arising in some theories with nonrenormalizable scalar potentials. Writing
\begin{equation}
\label{S1} \Phi(t,\textbf{r}) = e^{i\omega t} \phi(t,\textbf{r}),
\end{equation}
and varying the functional $\mathcal{E}$ [defined by Eq.~\eqref{functional1}] with respect to $\Phi$, we find the equation of motion for the field $\phi(t,\textbf{r})$:
\begin{equation}
\label{S2} -\ddot{\phi} - 2 i \omega \dot{\phi} = \widehat{\mathcal{D}} \phi,
\end{equation}
where $\widehat{\mathcal{D}}$ is the linear differential operator
\begin{equation}
\label{S3} \widehat{\mathcal{D}} = - \nabla^2 + \frac{\delta U}{\delta |\phi|^2} - \omega^2.
\end{equation}
Since Eq.~\eqref{S2} is linear in $\phi$ [due to the form of the potential $U$, see Eq.~\eqref{Uapproximate}], the evolution of small perturbations $\delta \phi$ about the background spinning supersymmetric Q-ball configurations is described by a similar equation:
\begin{equation}
\label{S4} - (\delta\phi)^{\cdot\cdot} - 2 i \omega (\delta\phi)^{\cdot} = \widehat{\mathcal{D}} \delta\phi.
\end{equation}
The solutions $\delta\phi_\lambda$ of the above equation are easily found:
\begin{equation}
\label{S5} \delta\phi_\lambda = e^{i \gamma t} \xi_\lambda,
\end{equation}
where $\gamma = -\omega \pm \sqrt{\omega^2 + \lambda}$, $\xi_\lambda$ and $\lambda$ being the eigenvectors and eigenvalues of $\widehat{\mathcal{D}}$: %
\begin{equation}
\label{S6} \widehat{\mathcal{D}} \xi_\lambda = \lambda \xi_\lambda.
\end{equation}
From Eq.~\eqref{S5}, we get that the background solution is unstable [i.e. $\delta\phi_\lambda(t)$ grows unboundedly with time] if there exists a $\lambda$ such that $\mbox{Im}[\gamma] < 0$. However, this is not the case since $\gamma$ is real. In fact, writing Eq.~\eqref{S6} as
\begin{equation}
\label{S7} -\nabla^2 \xi_\lambda = \left( -\frac{\delta U}{\delta |\phi|^2} + \omega^2 + \lambda \right) \! \xi_\lambda,
\end{equation}
and remembering that the eigenvalues of the operator $-\nabla^2$ are strictly positive real numbers, we obtain
\begin{equation}
\label{S8} \omega^2 + \lambda > \frac{\delta U}{\delta |\phi|^2} \geq 0,
\end{equation}
where in the last inequality we used Eq.~\eqref{Uapproximate}. The above equation shows that $\gamma$ is a real quantity, as anticipated.


\section{V. Conclusions}

We have succeeded in obtaining, analytically, nontopological solitonic solutions with nonvanishing angular momentum in (3+1)-dimensional Minkowski spacetime in the theory of a self-interacting complex scalar field carrying a conserved global $U(1)$ charge.

This kind of solitons (known as spinning Q-balls) naturally emerge in a particular class of supersymmetric extensions of the standard model of particle physics where supersymmetry is spontaneously broken at low energy. The scalar field is in this case a gauge-singlet combination of squarks and sleptons corresponding to some flat direction of the supersymmetric potential, while the conserved global charge is some combination of baryon and lepton numbers. In this class of models an effective potential for the flat directions arises due to the breaking of supersymmetry.

We have shown that such a type of potential admits, as the nonperturbative ground state of the theory, axisymmetric Q-balls whose angular momentum is directed along the axis of symmetry. Working in the limit of large charges, we have found that the state of a {\it spinning supersymmetric Q-ball} can be labeled by the triple $(Q,l,m)$, where $Q$ is the conserved $U(1)$ charge, $l$ is positive integer that can take the values $l=|m|$ and $l=|m|+1$ and defines the parity of the state, $P = (-1)^l$, while $m$ is a integer which gives the projection of the angular momentum on the axis of symmetry through $J = -mQ$.

Moreover, we have found the expressions for the energy and radius of spinning supersymmetric Q-balls, which fully determine their astrophysical and cosmological properties. It turns out that they do not explicitly depend on $m$ and, at fixed charge, are increasing functions of $l$. This indicates that spinning supersymmetric Q-balls are indeed excitations of spherically-symmetric supersymmetric Q-balls. They are classically stable --due to conservation of angular momentum and parity-- and stable against small perturbations about their classical configurations, and could form during collisions between supersymmetric Q-balls and/or during the process of fragmentation of the Affleck-Dine condensate at the end of inflation.

Finally, in the case $(l,m)=(0,0)$, we reobtain the well-known solution and properties of parity-even, supersymmetric Q-balls with vanishing angular momentum already analyzed in the literature.


\end{document}